\documentclass{article}

\usepackage{arxiv}

\usepackage[utf8]{inputenc} % allow utf-8 input
\usepackage[T1]{fontenc}    % use 8-bit T1 fonts
\usepackage{hyperref}       % hyperlinks
\usepackage{url}            % simple URL typesetting
\usepackage{booktabs}       % professional-quality tables
\usepackage{amsfonts}       % blackboard math symbols
\usepackage{nicefrac}       % compact symbols for 1/2, etc.
\usepackage{microtype}      % microtypography
\usepackage{lipsum}		% Can be removed after putting your text content

\usepackage{amsmath}
\usepackage{algorithm}
\usepackage[noend]{algpseudocode}
\usepackage{color, colortbl}
\definecolor{HColor}{gray}{0.9}
\usepackage{multirow}
\usepackage{graphicx}

\title{Competitive Influence Maximization: Integrating Budget Allocation and Seed Selection}

\date{} 					% Or removing it

\author{
  Amirhossein Ansari \\
  Shiraz University, Shiraz, Iran\\    
  \texttt{a.ansari@cse.shirazu.ac.ir} \\
  \And  
  Masoud Dadgar \\
  Shiraz University, Shiraz, Iran\\    
  \texttt{dadgar@cse.shirazu.ac.ir} \\
  \And  
  Ali Hamzeh \\
  Shiraz University, Shiraz, Iran \\    
  \texttt{ali@cse.shirazu.ac.ir} \\
  \And  
  J{\"o}rg Schl{\"o}tterer \\
  University of Passau, Passau, Germany \\    
  \texttt{joerg.schloetterer@uni-passau.de} \\
  \And  
  Michael Granitzer \\
  University of Passau, Passau, Germany \\    
  \texttt{michael.granitzer@uni-passau.de} \\  
}

\begin{document}
\maketitle

\begin{abstract}
Today, many companies take advantage of viral marketing to promote their new products, and since there are several competing companies in many markets, Competitive Influence Maximization has attracted much attention. Two categories of studies exist in the literature. First, studies that analyze the problem of which nodes from the network to select considering the existence of the opponents. Second, studies that focus on the problem of budget allocation. Although these studies have improved the problem in many aspects, their considered scenario is still incomplete. In this paper, we integrate these two lines of researches and propose a much more realistic scenario for the Competitive Influence Maximization problem. In our scenario, competition happens in two phases. First, parties identify the most influential nodes of the network. Then they compete over only these influential nodes by the amount of budget they allocate to each node. Also, despite the common assumption in all previous budget allocation studies that the action space is continuous, we consider the action space to be discrete. This assumption is more connected to real-world applications and significantly changes the problem. We model the scenario as a game and propose a novel framework for calculating the Nash equilibrium. Notably, building an efficient framework for this problem with such a huge action space and also handling the stochastic environment of influence maximization is very challenging. To tackle these difficulties, we devise a new payoff estimation method and a novel best response oracle to boost the efficiency of our framework.
\end{abstract}

% keywords can be removed

\section{Introduction}
\label{sec:intro}
Growing popularity of social networks like Instagram, Facebook, and their ability to propagate ideas and information so rapidly has made the Influence Maximization (IM) an eye-catching task. The IM problem in its classical form is to identify the top $k$ influential nodes of a social network that can influence (directly or indirectly) the largest number of nodes in the network~\cite{kempe2003maximizing}. Viral marketing is a proper application of IM. A company selects a seed set of customers and activate them using free products, hoping that product adoption would propagate in the network through the word of mouth effect~\cite{richardson2002mining}.

Kempe et al.~\cite{kempe2003maximizing} were the first to formulate the IM problem as a discrete stochastic optimization task and presented Independent Cascade (IC) and Linear Threshold (LT) models for influence diffusion. These stochastic diffusion models define how influence starts from the initial nodes and spreads through the network. They proved that under these two models, the IM problem is NP-hard and proposed a greedy seed selection algorithm that guarantees the $(1-1/e-\epsilon)$-approximation of the optimal solution.

In the real world, it is prevalent that two or more companies are competing in the same market, which means that the classical IM problem lacks a competitive extension~\cite{bharathi2007competitive}. In the Competitive Influence Maximization (CIM), several different companies are propagating their products or ideas simultaneously in the network. Also, the goal of each company is to activate the largest number of nodes and defeat the opponents. Based on the assumptions and scenarios considered in previous studies, they can be categorized into two subcategories:
\begin{enumerate}
  \item Studies that focus on the problem of which nodes from the network to select considering the existence of other opponents. In these studies, a competitive extension to the IC or LT models is used for diffusion dynamics. Among these studies, two scenarios are very famous. First, the follower's perspective scenario~\cite{bharathi2007competitive,carnes2007maximizing,bozorgi2017community,li2018dominated,yan2018minimum}, which supposes that seed selection happens in turn, and solves the problem by considering itself as the last player to commit in seed selection. So, it takes advantage of knowing the other opponents' seed sets. Second, the scenario that considers that seed selection happens simultaneously, and the main objective is to propose a framework for selecting the best method from a set of seed selection algorithms available \cite{lin2015learning,li2015getreal,ali2018boosting,ali2018novel}.

  \item Studies that address the budget allocation scenario~\cite{masucci2014strategic,masucci2017advertising,varma2018marketing,varma2019allocating}. In this scenario, companies can allocate different values of their budget to a node in the network, and nodes will prefer the product of the company that has offered a higher value. So, here, companies compete with each other by the amount of budget they allocate to each node in the network. The common assumption in all these studies is that the amount of budget allocated to a node is a real value. Also, the voter model is applied for influence dynamics.
\end{enumerate}

The seed selection problem (the first category) has long been studied in the literature. However, only a few works have analyzed the budget allocation scenario (the second category). We started our work by conducting an in-depth analysis of the seed selection studies (Section~\ref{sec:discussion_seed_selection}). We observed that among these studies, the scenario that seed selection happens at the same time is more realistic. However, even this scenario is still incomplete. As we will discuss, in this scenario, the most promising choice for parties is one of the effective non-competitive seed selection methods, and these methods largely agree on which nodes from the network to select. Therefore, the result of this scenario would be the state that parties are targeting the same nodes from the network, which shows that this modeling is still incomplete.

The studies in the budget allocation scenario also have the defect that the considered action space is continuous, meaning that the amount of the budget allocated to a node is a real value. In reality, we often see offers like 10\%, 15\% of discount. Offers like 12.22\% and 12.77\% are very rare. Also, a common realization of viral marketing is the example that companies offer free products with different values to influential customers. In this case, the company must decide which product to offer to which influential customer, and again the action space is discrete. These observations indicate that the value of the budget allocated to a node must be discrete instead of being a real value. Furthermore, when the action space is continuous, the whole network can be targetted by allocating a very small fraction of the budget to each node in the network. However, again, this consideration is far from reality.

\begin{figure}
  \centering
  \includegraphics[width=0.8\textwidth]{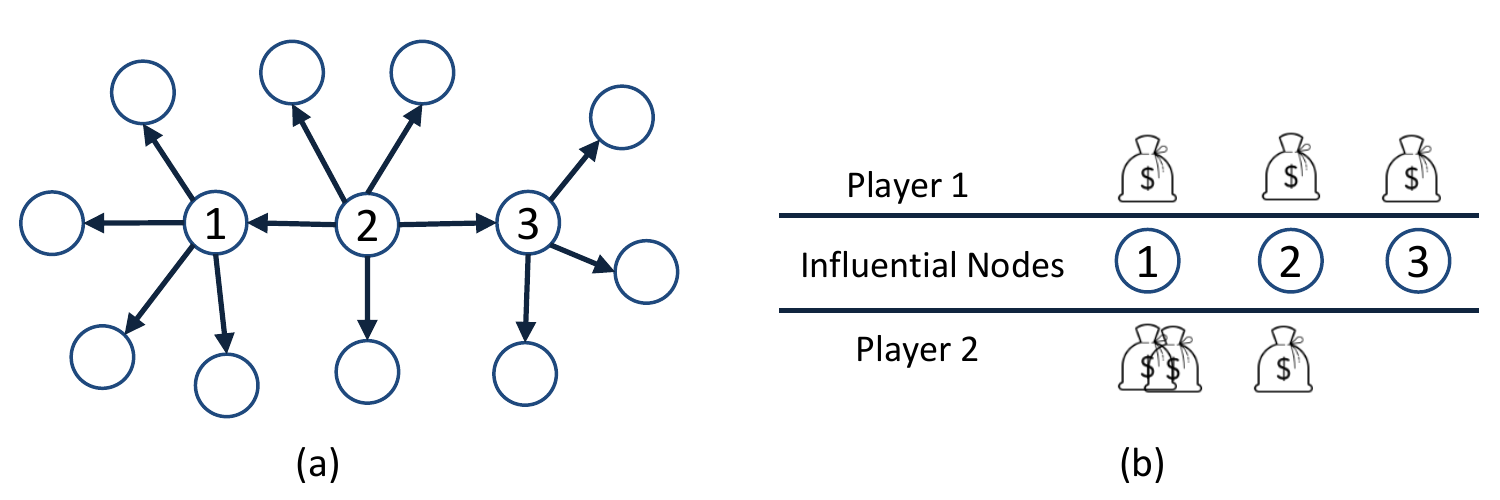}
\caption{An example of (a) a social network and (b) the proposed scenario}
\label{fig:example}
\end{figure}

The observations from analyzing the seed selection studies and also the fact that the action space in the budget allocation scenario tends to be discrete, motivated us to integrate seed selection and budget allocation to propose a new scenario for the CIM problem. In our proposed scenario, the parties first identify the influential nodes of the network. Then, they compete over only these nodes, not the whole network, by the value of the budget they allocate to each influential node. Also, the considered action space is discrete. This modeling of the problem improves the discussed issues with previous studies. 
Figure~\ref{fig:example} demonstrates a simple example of the proposed scenario. Both parties first recognize that nodes 1, 2, and 3 are the most influential nodes in the given social network. They have three units of budget and compete with each other over these influential nodes. Player 1 decides to allocate one unit of budget (a package that costs one unit of budget) to each of these influential nodes. However, player 2 decides to allocate two units to node 1 and one unit to node 2. In this case, node 1, for example, will prefer player 2 as he has suggested a package with a higher value.

We use game theory to analyze the scenario and calculate the Nash equilibrium by the Double Oracle algorithm~\cite{mcmahan2003planning}. This algorithm is specially designed for two-player zero-sum games with large action spaces like ours. For the diffusion dynamics, we have applied the competitive Independent Cascade (competitive-IC) model. Estimating the payoff in this diffusion model is very challenging. It is proven that calculating the influence spread under the IC model is \#P-hard~\cite{chen2010scalable}. To tackle this difficulty, we have designed a novel method that assigns an influential value to nodes based on the concept of Reverse Reachable (RR) sets~\cite{borgs2014maximizing}. Also, our method is applicable in LT and Trigerring based diffusion models, since the concept of RR-sets is generalizable to these models \cite{tang2014influence}.
To list our contributions:
\begin{itemize}
\item We propose the two-phase budget allocation scenario where seed selection and budget allocation are integrated. This scenario improves previous works on seed selection by changing the focus of the problem to convincing influential nodes to act as seed rather than which nodes from the network to select (we have a strong motivation for this change presented in Section~\ref{sec:discussion_seed_selection}).  Also, the budget allocation scenario is enhanced in the following ways: 1. The focus of the competition is reduced to only influential nodes instead of the whole network. 2. The action space is considered to be discrete. 3. The scenario is extended to the IC, LT, and Triggering based diffusion models.
\item We develop a novel competitive aware budget allocation framework for the CIM problem. We especially design a new payoff estimation method and a best response oracle that significantly improves the efficiency of our framework.
\item We conduct exhaustive experiments on real-world networks to evaluate different aspects of our work. This is the first time that the budget allocation scenario is tested on real-world networks. The evaluations of \cite{masucci2017advertising,varma2018marketing,varma2019allocating} consist of only simple simulations.
\end{itemize}

This paper is organized as follows: Section~\ref{sec:realted} reviews the related work.  In Section~\ref{sec:preliminaries}, some preliminaries are described. In Section~\ref{sec:discussion_seed_selection}, our discussion on seed selection studies is presented, and we discuss why these studies are unrealistic. The proposed scenario and the proposed framework will be described in Section~\ref{sec:scenario} and Section~\ref{sec:framework}. Experiments are reported in Section~\ref{sec:experiments}, and finally, we conclude our work in Section~\ref{sec:conclusion}.

\section{Related Work}
\label{sec:realted}
Carnes et al.~\cite{carnes2007maximizing} and Bharathi et al.~\cite{bharathi2007competitive} are among the firsts that have addressed the Competitive Influence Maximization (CIM) problem. They first provide extensions to the Independent Cascade (IC) model for the competitive environment, and then analyzed the problem given the opponent's seed set, the follower's perspective. They similarly showed that under their diffusion models, the greedy algorithm provides the $(1-1/e-\epsilon)$-approximation for the second player (the last player). Bharathi et al. further studied the best response strategy for the first player in very simple cases, like when the graph is directed lines.

Lin et al.~\cite{lin2015learning} considered the scenario that seed selection happens in a multi-round manner and proposed a framework based on Q-learning to learn the optimal strategy for the parties. In their framework, actions were considered as single-party influence maximization algorithms. Also, diffusion happens according to the competitive Linear Threshold (competitive-LT) model, and the reward was calculated in the last round by difference of the number of activated nodes by parties. Their framework is applicable when the opponents' strategy is known or unknown. 
Li et al.~\cite{li2015getreal} considered a more realistic scenario where parties select their seed sets simultaneously. They modeled the problem as a game with seed selection algorithms as actions and the expected influence as payoffs. Also, diffusion happens according to the competitive-IC model. These two works do not propose a new strategy for seed selection. Instead, a framework for selecting the best strategy from a set of seed selection algorithms available is proposed. Their solution is more practical since they do not consider any information about the seed set of the opponent.

In \cite{bozorgi2017community}, a new model based on the competitive-LT model where each node can think about the incoming influence, and an influence maximization algorithm based on community detection were proposed. In \cite{li2018dominated}, time limitation and time delay in the propagation process have been taken into account, and the competitive-IC model with meeting events was proposed. In \cite{yan2018minimum}, given the opponents' seed sets, they have tried to select the seed set that its spread will pass a certain threshold with a minimum cost. All these works have solved the problem with the assumption that the opponent's seed set is given (from the follower's perspective).

In \cite{ben2018spread}, a two-player zero-sum extensive-form game is proposed that simulates the idea of the CIM problem. The players have an equal number of tokens, and in each stage of the game, they should choose one node from the graph to put a token on. When the number of tokens on a node reaches a threshold, the node will become activated and spreads its tokens to the neighbors. They have used Alpha-Beta pruning and Monte Carlo tree search algorithms to find the optimal strategy for the players.

In \cite{ali2018boosting} and \cite{ali2018novel}, the Q-learning based framework proposed in~\cite{lin2015learning} is improved in the following way. In \cite{ali2018boosting}, transfer learning is applied to avoid retraining the Q-learning method when the social network changes. In \cite{ali2018novel}, the factor of time is integrated into the competition, and a nested Q-learning algorithm is proposed.

Masucci and Silva~\cite{masucci2014strategic} are the first to introduce the budget allocation scenario. They analyzed the case of two competitors and used game theory, especially Colonel Blotto games, to calculate the Nash equilibrium. They also used the voter model with discrete states for their diffusion dynamics. Later in~\cite{masucci2017advertising}, they extended the problem to the case of more than two players and presented the voter model with ranking scores. In both works, the considered action space was continuous, meaning that any fraction of the budget could be allocated to a node. Also, all players had an equal amount of budget.

Varma et al.~\cite{varma2018marketing} extended the voter model to the case where nodes have continuous states representing their opinion toward the players. Each player's objective is to maximize the overall opinion in the network. They mainly considered the case of two players, and again, the action space was continuous, and also the players' amount of budget could be different. They showed that under their proposed model, the game has a pure Nash equilibrium. In~\cite{varma2019allocating}, they further extended their work to the case that marketing campaigns are repeated.

\section{Preliminaries}
\label{sec:preliminaries}
\subsection{Competitive Independent Cascade Model}
\label{sec:competitive_ic}
The competitive Independent Cascade (competitive-IC) model proposed in~\cite{bharathi2007competitive} is a widely accepted diffusion model for the competitive environment. A directed graph $G = (V,E)$ is given and an activation probability $P_e\in [0,1]$ is assigned to each edge. Nodes have two states, they are either $\mathit{inactive}$ or $\mathit{activated_i}$. In the latter case, $i=1,2,\dots,n$ denotes the player, for whom the node is activated. Consider that $S_i$ denotes the seed set of the player $i$. Given the seed sets, influence propagates in the network through a progressive process as follows:
\begin{itemize}
\item In the first step, nodes in each set $S_i$ become activated and their states are set to $\mathit{activated_i}$.
\item In step $t$, in a random order, nodes activated in step $t-1$ will try to activate their free neighbors. The activation succeeds with probability $P_e$, and newly activated nodes take the state of the node who has activated them.
\item The process terminates when no new activation occurs.
\end{itemize}
According to the process, each newly activated node has only a single chance to activate its free neighbors, and also active nodes never change state. The influence spread of player $i$, $I(S_i)$, denotes the number of nodes activated by $S_i$ when diffusion terminates.

\subsection{Reverse Reachable Sets}
\label{sec:rr_sets}
The Reverse Reachable (RR) sets~\cite{borgs2014maximizing} is a proven method to approximate the influence spread and avoid the limitations of the greedy algorithm. In this section, we first describe this concept in the non-competitive environment, then it is described how this method is extended to the competitive problem when the opponent's seed set is given.

\subsubsection{RR-sets in Non-competitive Problem}
\label{sec:non_competitive_rr_sets}
This is the definition of the random RR-sets under the classical IC model~\cite{borgs2014maximizing,tang2014influence}:
\begin{quote}
  Let $g$ be a graph in which diffusion is deterministic, meaning that, if there is a path from a node $u$ to a node $v$, then $u$ can influence $v$. An RR-set for a node $v$ in the deterministic graph $g$ is a set that includes all the nodes in $g$ that has a path to $v$. $v$ is called the root node. A random RR-set is an RR-set created for a root node selected randomly, also the deterministic graph $g$ is sampled by removing each edge of $G$ with probability $1-P_e$ (the root node and deterministic graph $g$ are both sampled randomly).
\end{quote}

Intuitively, the nodes in an RR-set have a chance to activate the root node of the RR-set, if they are selected as the seed set. Algorithm~\ref{alg:rr_set} describes how an RR set is created algorithmically. First, the RR-set and a queue are initialized with the root node selected uniformly at random (lines 1 and 2). Then, we start to move backward from the root node and add each visiting node to the RR-set according to the probability assigned to the corresponding edge. The process of moving backward is repeated for each node newly added (lines 3 to 6)~\cite{tang2014influence}.

\begin{algorithm}  
  \caption{Random RR set creation algorithm}
  \label{alg:rr_set}
  \begin{algorithmic}[1]      
    \State Choose node $v$ from $G$ uniformly at random
    \State $RS \gets \{v\}, Q \gets \{v\}$
    \While{$Q$ is not empty}
      \State $u \gets Q.pop()$
      \ForAll{$w \in \mathit{in\_neighbors}(u)$}
        \State add $w$ to $Q$ and $RS$ according to the probability $P_{w \rightarrow u}$
      \EndFor
    \EndWhile
    \State \textbf{return} $RS$      
  \end{algorithmic}
\end{algorithm}

RR-set based methods like TIM/TIM+~\cite{tang2014influence} and IMM~\cite{tang2015influence} build their seed selection algorithm upon this idea that when a sufficient number of random RR-sets are created, the expected spread of any seed set $S$ can be estimated by the fraction of the RR-sets covered by the seed set (A seed set covers an RR set if it has a node in that):
\begin{equation}
  \label{eq:non_cim_rr_set}
  E[I(S)] = \frac{N}{\theta}C(S)
\end{equation}
$N$ is the number of nodes, $\theta$ is the number of RR-sets created, and $C(S)$ is the number of RR-sets covered by the seed set $S$~\cite{tang2014influence,tang2015influence}. Algorithm~\ref{alg:tim} describes the outline of the RR-set based seed selection methods. First, a sufficient number of random RR-sets are created (line 1). Then, iteratively, the seed set is filled with the node that covers the largest number of the RR-sets (lines 3 to 6). Note that in each iteration, the RR-sets covered by the selected node are removed (line 6).

\begin{algorithm}  
  \caption{Outline of the RR-set based seed selection algorithms (TIM+\cite{tang2014influence}, IMM\cite{tang2015influence})}
  \label{alg:tim}
  \begin{algorithmic}[1]      
    \State $\mathcal{R} \gets$ sufficient number of random RR-sets are created    
    \State $S\gets \{\}$      
    \For{$i=1$ to $k$}
      \State $v \gets$ the node that covers the largest number of RR-sets in $\mathcal{R}$
      \State $S \gets S \cup \{v\}$
      \State remove the RR-sets covered by $v$ from $\mathcal{R}$
    \EndFor    
    \State \textbf{return} $S$      
  \end{algorithmic}
\end{algorithm}

\subsubsection{RR-sets in Competitive Problem}
\label{sec:competitive_rr_sets}
Lu et al.~\cite{lu2015competition} extended the definition of RR-sets to the general case where two items that complement or compete with each other are diffused in the network. By this extension, they proposed the general-TIM algorithm for seed selection with the assumption that the other player's seed set is given. They proved that, in specific conditions, their algorithm provides the $(1-1/e-\epsilon)$ approximation with a high probability. They further show that in the pure competition (our problem) their conditions completely hold.

According to their general definition of RR-sets, for creating a random RR-set for the competitive problem (under the competitive-IC model), the process is entirely the same as the non-competitive process, except that the backward phase stops when a node from the opponent's seed set is visited. We refer to the general-TIM algorithm with competitive RR-sets as the competitive-TIM algorithm.

\section{Discussion on Seed Selection Scenarios}
\label{sec:discussion_seed_selection}
In this section, we conduct a step by step investigation regarding the scenarios where the focus of the competition is on seed selection, and discuss why these modelings are not realistic. For simplicity, our analysis is mainly around the case of two players, and the competitive-IC model is used for diffusion.

First, consider the follower's perspective scenario. In this scenario, the second player selects his seed set after the first player, and as a result, has complete knowledge of the opponent's seed set. For this player, the greedy algorithm provides a solution with the $(1-1/e-\epsilon)$-approximation of the best response~\cite{bharathi2007competitive}. As described in Section~\ref{sec:competitive_rr_sets}, the competitive-TIM provides the same approximation with a high probability and is much more efficient than the greedy method~\cite{lu2015competition}. Therefore, competitive-TIM is a perfect choice for the second player. Now consider the first player. As far as we know, no practical choice has been proposed for the first player, and we should refer to the non-competitive seed selection algorithms. There are generally two categories of seed selection strategies in the non-competitive literature. First, methods that do not guarantee any approximation. Second, algorithms with an approximation guarantee $(1-1/e-\epsilon)$, like CELF~\cite{leskovec2007cost}, CELF++~\cite{goyal2011celf++}, TIM+~\cite{tang2014influence} and IMM~\cite{tang2015influence}. It makes much more sense to consider that the first player selects his seed set by one of the algorithms from the second category.

\begin{figure}
  \centering
  \includegraphics[width=0.65\textwidth]{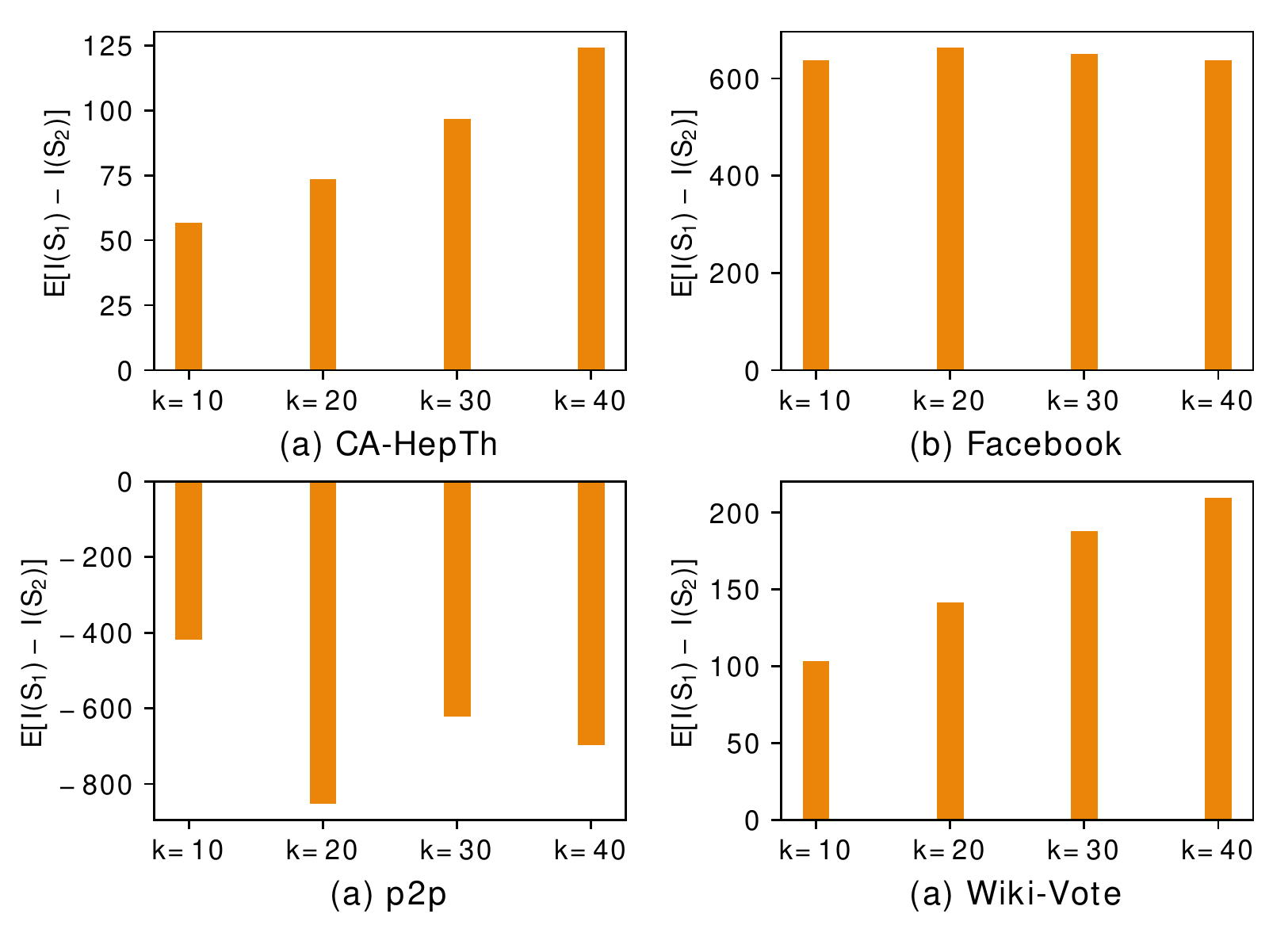}
\caption{$E[I(S_1)-I(S_2)]$ in the competition where the 1\textsuperscript{st} player adopts TIM+ and the 2\textsuperscript{nd} player adopts competitive-TIM}
\label{fig:tim_vs_comp_tim}
\end{figure}

We conduct an experiment where the first player adopts the TIM+ algorithm (a representative from the second category), and the second player uses the competitive-TIM. Also, for both players, the seed set size is equal to $k$ ($|S_1|=|S_2|=k$). The diffusion happens according to the competitive-IC model and is repeated 5000 times to get an exact estimate of $E[I(S_1)-I(S_2)]$. Note that $I(S_i)$ denotes the number of nodes activated by $S_i$. Results are reported in Figure~\ref{fig:tim_vs_comp_tim}~\footnote{CA-HepTh, Facebook, p2p, and Wiki-Vote are the four datasets used in our experiments. Their statistics are described in Section~\ref{sec:experiments_settings}}. 
When the value of $E[I(S_1)-I(S_2)]$ is positive, the first player has caused more influence spread and won the game. Similarly, when this value is negative, the second player has won the game. Results show that the second player has lost the game in 3 out of 4 datasets, although he had complete knowledge of the opponent's seed set. Therefore, it seems that the first player has more chance to win, and the considered scenario is biased.

Now, consider the scenario where the objective is to propose a framework for selecting the best seed selection strategy. To analyze this case, we consider the scenario proposed in \cite{li2015getreal}. In this scenario, seed selection happens at the same time, and also seed sets can overlap. Overlapping nodes will be assigned to one of the players randomly. Considering the previous discussion, and also the fact that there is no competitive seed selection algorithm specific to this case, we would get again to the non-competitive seed selection methods with the greedy approximation ratio as the best choices. 
Note that the main difference among these methods is their running time, not their effectivity. Therefore, it is sensible that both players use IMM~\cite{tang2015influence}, which is the most efficient and scalable algorithm from this category, and so, their seed sets would be entirely the same. Even if the players adopt different methods from this category that has an approximation guarantee, again they will almost select the same nodes. 
To show this, we have calculated the number of nodes that overlap between the seed sets selected by CELF~\cite{leskovec2007cost}, CELF++~\cite{goyal2011celf++}, TIM+~\cite{tang2014influence} and IMM~\cite{tang2015influence} algorithms and report the result in Table~\ref{tab:overlapping_nodes}~\footnote{For CELF/CELF++ we have used the code in \url{https://github.com/jjboo/InfMax}, and for TIM+ the code in \url{https://sourceforge.net/projects/timplus/}, and for IMM the code in \url{https://sourceforge.net/projects/im-imm/}. Also the parameters of these experiments are: 20000 MC simulations in CELF/CELF++ and $ \epsilon=0.05 $ in TIM+/IMM.}. 
The number of overlappings will even increase if we reduce the approximation error by increasing the number of Monte Carlo simulations in CELF and CELF++ and decrease $ \epsilon $ in TIM+ and IMM. Therefore, the result of this scenario is the state that the players are targeting the same initial adopters. In conclusion, this observation indicates that the competition is beyond the choice of the seed selection method.

\begin{table}
\caption{$|S_1 \cap S_2 \cap S_3 \cap S_4|$ where $S_1$ is identified by CELF~\cite{leskovec2007cost}, $S_2$ by CELF++~\cite{goyal2011celf++}, $S_3$ by TIM+~\cite{tang2014influence} and $S_4$ by IMM~\cite{tang2015influence}}
\label{tab:overlapping_nodes}
\centering
\begin{tabular}{ccccc}
\toprule
& $k=10$ & $k=20$ & $k=30$ & $k=40$\\
\midrule
CA-HepTh  & 9  & 16 & 24 & 29\\
Facebook  & 10 & 17 & 17 & 25\\
p2p       & 10 & 17 & 24 & 34\\
Wiki-Vote & 10 & 18 & 28 & 37\\
\bottomrule
\end{tabular}
\end{table}

\section{Proposed Scenario}
\label{sec:scenario}
Suppose that different companies are competing for more product sales in a social network. Each company has a marketing budget and several advertising packages with different values. The companies first identify the most influential nodes of the network, and then they decide which package should be allocated to which influential node. Nodes will prefer the package with the highest value.

Now we formally define the scenario. Consider that the set $S=\{s_1,s_2,...,s_n\}$ with size $n$ denotes the set of most influential nodes. Also, suppose that player $i$ has a marketing budget $k_i$ and his available advertising packages are $D_i=\{d_1, d_2,...,d_l\}$, where $d_i$ denotes the value of the package, or equivalently the cost of the package for the company. By this formulation, the actions available to player $i$ would be $A_i = \{a_j|a_j \in \{0\}\cup D_i \}$, where $a_j$ indicates the value of the package allocated to the $j$th node in $S$, and also $\sum_{j=1}^{n} a_j \leq k_i$. If $a_j=0$, it means that the player has ignored the node. Players simultaneously select their actions. Each node $s_i$ will be assigned to the seed set of the player that has offered the node a package with higher value compared to the other players. If two or more players have allocated the highest value, the node will be assigned to one of them randomly. When players play their actions (at the same time), their seed sets, $S_i$, can be identified, and then diffusion happens according to the competitive-IC model described in Section~\ref{sec:competitive_ic}.

As an example, consider Figure~\ref{fig:example} part b. In this example, both players have an equal amount of budget, $k_1=k_2=3$. Also, they have packages with similar values, $D_1=D_2=\{1,2,3\}$. The set of most influential nodes is $S=\{s_1,s_2,s_3\}$. The first player has adopted action $a^1=\{1,1,1\}$, and the second player has adopted action $a^2=\{2,1,0\}$. Therefore, node $s_1$ will be assigned to the second player (who has offered a higher value) and node $s_3$ to the first player. Also, node $s_2$ would be assigned to one of the players randomly, as both players have allocated a package with equal value to this node. Suppose that the result is that this node is assigned to the second player. Finally, the seed sets would be $S_1=\{s_3\}$ and $S_2=\{s_1,s_2\}$. Then, diffusion happens, and the winner of the game would be identified, which is player 2 in this example.

\section{Proposed Framework}
\label{sec:framework}
We model the scenario as a normal form game and propose a novel framework for the case of two players. As the first step, $n$ top influential nodes of the network (the nodes in set $S$, $|S|=n$) are identified using the TIM+ algorithm described in Section~\ref{sec:non_competitive_rr_sets}. This is the first phase of our framework. We define the utility (payoff) for the first player as $E[I(S_1)-I(S_2)]$, where $I(S_1)$ and $I(S_2)$ are the number of nodes activated by the first player's seed set ($S_1$) and the second player's seed set ($S_2$). Respectively, the second player's payoff would be $E[I(S_2)-I(S_1)]$. In this way, we have formed a zero-sum game, $\mathit{payoff}_1+\mathit{payoff}_2=0$. Zeros-sum games model the case of pure competition where any gain by one player results in the same amount of loss for the other player.

In a multi-agent environment, players can follow two kinds of strategies. Selecting one single action and playing it, pure strategy, or playing according to a probability distribution assigned to actions, mixed strategy. The pure strategy is a mixed strategy where probability one is assigned to a specific action. The most common and prominent solution concept in normal-form games is the Nash equilibrium. Nash equilibrium is a stable point where each player is playing the best response against its competitor. Therefore, no rational player would like to deviate from the strategy playing in Nash even if he gets to know the opponent's strategy. Every finite game has at least one Nash equilibrium~\cite{shoham2008mas}.

Consider that the value of the game is defined as the first player's payoff in equilibrium (the state where both players are playing their respective Nash strategy). According to the Minimax theorem~\cite{shoham2008mas}, in zero-sum games, the Nash strategy guarantees the player playing this strategy a minimum amount of payoff, which is equal to the value of the game. Note that when both players have an equal amount of budget, $k_1=k_2$, and also both have packages with the same values, $D_1=D_2$, the game would be symmetric. In symmetric games, the value of the game is zero~\cite{peters2015game}. Thus, in this case, the Nash strategy guarantees a payoff of at least zero, or in other words, guarantees not to lose any competition. This is a very promising guarantee.

To calculate the Nash equilibrium, we faced two challenging difficulties:
\begin{enumerate}
  \item Our action space is considerably large. As an example, for the parameters $k_1=k_2=20$, $D_1=D_2=\{1,2,...,20\}$ and $n=25$, size of the action space is 1,761,039,350,070.
  \item Considering the size of our action space, an efficient method for approximating the value of payoff, $E[I(S_1)-I(S_2)]$, is required. The widely adopted solution of directly diffusing the seed sets in the network~\cite{li2015getreal,lin2015learning,ali2018boosting,ali2018novel} is not applicable here. Since, in this method, to reach an acceptable approximation, the diffusion must be repeated many times, which takes a long time.
\end{enumerate}

To tackle the first difficulty, we have used the Double Oracle algorithm proposed in ~\cite{mcmahan2003planning} for two-player zero-sum games. The general idea of the method is that it considers a restricted game, where only a subset of the actions is available to each player. Then, the action space of the restricted game is iteratively enlarged until the algorithm converges. Algorithm~\ref{alg:double_oracle} describes the outline of the Double Oracle algorithm. The available action sets are initialized with an arbitrary action from the action space (line 1). In each iteration, the Nash equilibrium of the restricted game is calculated (line 3). Using a best response oracle, the best response in pure strategies against the calculated equilibriums are identified and added to the action sets (lines 4-8). The algorithm converges when the calculated best responses are already in the action sets: both $a_1$ and $a_2$ are in $A'_1$ and $A'_2$. In this case, $nash_1$ and $nash_2$ are the exact equilibrium of the game with the whole action sets. Another termination condition is to continue until $v_1 - v_2 < \epsilon$, in this way, an approximation is calculated~\cite{mcmahan2003planning}. The algorithm guarantees to converge, and practically since only best responses are added in each iteration, it converges much faster than directly calculating the Nash equilibrium~\cite{mcmahan2003planning}.

The Double Oracle algorithm works best in the problems that an efficient oracle for calculating the pure best response exists. To propose a very fast best response oracle and solve the problem of payoff estimation, we reuse the RR-sets created in the first phase and develop a heuristic method that assigns an influential value to the nodes in $S$. Then, we approximate the value of payoff using these calculated values. Suppose that the adopted actions by the first player and the second player are respectively $a^1=\{a_1^1, a_2^1, ..., a_n^1\}$, and $a^2=\{a_1^2, a_2^2, ..., a_n^2\}$, the payoff for the first player would be:
\begin{equation}
  \label{eq:payoff}
    \mathit{payoff}_1(a^1, a^2) =\!\! \sum_{s_i \in S_1}\!\!\!\mathit{value}(s_i) -\!\! \sum_{s_i \in S_2}\!\!\!\mathit{value}(s_i) = \sum_{i=1}^{n} f(a_i^1,a_i^2)
\end{equation}
where the function $f(.)$ denotes the gain by node $s_i$ in the first player's payoff and is defined as:
\begin{equation}
  \label{eq:payoff_f}
  f(a_i^1, a_i^2) =
  \begin{cases}
    0 & \text{if}\; a_i^1 = a_i^2 \\
    -\mathit{value}(s_i) & \text{if}\; a_i^1 < a_i^2 \\    
    +\mathit{value}(s_i) & \text{if}\;  a_i^1 > a_i^2 \\
  \end{cases}
\end{equation}
The above equation comes directly from the scenario and is common in other budget allocation studies~\cite{masucci2014strategic,masucci2017advertising}. The assignments in function $f(.)$ is almost clear. When the player wins the node by assigning a higher value so his payoff would increase by $\mathit{value}(s_i)$, and when he loses the node, his payoff will decrease. Also, when both players ignore the node ($a_i^1=a_i^2=0$), the gain from this node for both players would be zero. In addition, when $0<a_i^1=a_i^2$ again, the gain would be zero. Since, according to the scenario, in this case, the node will be assigned to one of the players randomly, and as our game is zero-sum, the expected gain is zero for both players. Calculation of $\mathit{payoff}_2$ is straight forward.

Note that the Nash strategy guaranteed a minimum amount of payoff, which is zero in the symmetric case. However, we are not calculating the exact value of payoff and the error of our heuristic influential value estimation method decreases the guaranteed value. Our experiments will show that this error is too small to affect the performance of the framework. Also, when the second termination condition is used in the double oracle algorithm, again, the guaranteed payoff would decrease by the value of the $\epsilon$ parameter.

In summary, our framework works in the following steps:
\begin{enumerate}
  \item The set of most influential nodes, $S$, is identified by TIM+.
  \item The influential values of the nodes in $S$ are estimated.
  \item The Nash equilibrium is calculated using the Double Oracle algorithm.
\end{enumerate}
Next we will describe our heuritic influential value estimation method and the best response oracle.

\begin{algorithm}  
  \caption{The Double Oracle algorithm~\cite{mcmahan2003planning}}
  \label{alg:double_oracle}
  \begin{algorithmic}[1] 
    \State Initialize $A'_1$ and $A'_2$ with an arbitrary action from $A_1, A_2$
    \Repeat
      \State $nash_1, nash_2 \gets$ Calculate the nash equilibrium of the game where the first player can only play the actions in $A'_1$ and the second player can only play the actions in $A'_2$
      \State $a_1 \gets \mathit{BestResponseOracle}(nash_2)$
      \State $v_1 \gets \mathit{payoff}_{1}(a_1, nash_2)$
      \State $a_2 \gets \mathit{BestResponseOracle}(nash_1)$
      \State $v_2 \gets \mathit{payoff}_{2}(nash_2, a_2)$
      \State Add $a_1$ to $A'_1$, and Add $a_2$ to $A'_2$
    \Until{The termination condition is satisfied}
  \end{algorithmic}
\end{algorithm}

\subsection{Influential Value Estimation Method}
\label{sec:value_estimation}
Here we describe how we have reused the RR-sets created by the TIM+ algorithm in the first phase of our framework to assign an influential value to the nodes in $S$. In the TIM+ algorithm, $\theta$ random RR-sets are created according to the non-competitive definition of RR-sets, definition 1. Consider that $\mathcal{R}$ denotes this set of RR-sets. Recall from Section~\ref{sec:non_competitive_rr_sets}, that an RR-set is a set of nodes that have a chance to activate the root node of the RR-set. Also, the influence spread of any seed set $S'$ can be estimated by $\frac{N}{\theta}C(S')$, where $N$ is the number of nodes in the network and $C(S')$ is the number of RR-sets that the seed set has at least a node in them (refer to Equation~\ref{eq:non_cim_rr_set}).

Now, we describe our extension to the RR-set based methods. A node covers an RR-set if it is a member of the RR-set, and so has a chance to activate the root node of the set. In our problem, each node $s_i \in S$ can have three states. It is either in $S_1$ (belongs to the first player), or in $S_2$ (belongs to the second player), or it is ignored by the players and no budget is allocated to it (it is like the nodes out of $S$). Considering these cases, the effect of covering an RR-set that is also covered by some other nodes in $S$ would not be equal to covering an RR-set that is just covered by a single node. This is the main difference between our problem and previous problems. To apply this difference, we assign a weight to each RR-set $r \in \mathcal{R}$: $w(r) = 1/|S \cap r|$, where $|S \cap r|$ indicates how many nodes from $S$ cover $r$ (or equivalently, number of the nodes in $S$ that are also in $r$). Suppose that $T(s \in r)$ returns 1 when $s \in r$ is true and 0 otherwise. We estimate the influential value of a node $s \in S$ by the sum of the weights of the RR-sets it covers:
\begin{equation}
  \textstyle
  \mathit{value}(s) = \frac{N}{\theta} \sum_{r \in \mathcal{R}} T(s \in r).w(r)
\end{equation}

Note that when the whole $S$ belongs to one player, for example, when $S_1=S$ and $S_2=\{\}$, our method's estimated value of the expected spread is exactly the same as the value estimated by non-competitive RR-set based methods (Section~\ref{sec:non_competitive_rr_sets}).  Also, the competitive RR-set based method (described in Section~\ref{sec:competitive_rr_sets}) is not applicable here, since they estimate the spread given the other player's seed set. However, we do not have any information about the other player's seed set.

\subsection{Best Response Oracle}
\label{sec:best_response_oracle}
The oracle calculates the best response in pure strategies against a given mixed strategy, $q$. The mixed strategy can be represented by the set of actions that probability greater than zero is assigned to them (support of the mixed strategy): $q=\{p_1:a^1, p_2:a^2,...,p_m:a^m\}$, where $p_i>0$ is the probability that the player would play $a^i$, and $a^1, a^2, ..., a^m$ are the actions in the support of $q$. Without loss of generality, we assume that the given mixed strategy is played by the second player, and we, as the first player, are searching for the best response. So, in this section, the notation of $\mathit{payoff}(,\!)$ indicates the payoff for the first player. Also, we use $k$ to denote the budget and $D$ to denote the set of the advertising packages of the player for whom we are trying to find the best response.

Payoff of adopting a pure strategy $B'$ against the given mixed strategy $q=\{p_1:a^1, p_2:a^2,...,p_m:a^m\}$ would be:
\begin{equation}
  \label{eq:q_payoff}
  \mathit{payoff}(B', q) = \sum_{i=1}^{m} \; p_i . \mathit{payoff}(B', a^i)
\end{equation}
where $\mathit{payoff}(B', a^i)$ is the payoff of $B'$ against the pure strategy (action) $a^i$ in the support of $q$. 
So, the oracle searches in pure strategies for the strategy $B$, that:
\begin{equation}
  \label{eq:best_resp}
  B = \underset{B'}{\operatorname{arg\,max}} \;\: \mathit{payoff}(B', q) = 
  \underset{B'}{\operatorname{arg\,max}} \;\: \sum_{i=1}^{m} \; p_i . \mathit{payoff}(B', a^i)
\end{equation}

Now we describe the algorithm to find the pure best response $B$. By our definition of actions, $B=\{b_1,b_2,...,b_n\}$ and each $a^i$ in the support of $q$ is $a^i=\{a_1^i, a_2^i, ..., a_n^i\}$, where each $b_j$, $a_j^i$ is the value of the package allocated to the $j$th node in $S$ under $B$ and $a^i$. According to Equation~\ref{eq:q_payoff} and Equation~\ref{eq:payoff}, the payoff of pure best response $B$ gainst the given mixed strategy $q$ would be:
\begin{equation}
  \mathit{payoff}(B, q) = \sum_{i=1}^{m} \; p_i . \mathit{payoff}(B, a^i) = \sum_{i=1}^{m} \sum_{j=1}^{n} \; p_i . f(b_j, a_j^i)
\end{equation}
and if we define the function $F_j(b_j)$ in the following way:
\begin{equation}
  F_j(b_j) = \sum_{i=1}^{m}\;p_i \: . \: f(b_j, a_j^i)
\end{equation}
Then, the payoff can be calculated based on the value of each $b_j$:
\begin{equation}  
  \mathit{payoff}(B, q) = \sum_{j=1}^{n} \left[ \sum_{i=1}^{m}  \; p_i . f(b_j, a_j^i) \right] = \sum_{j=1}^{n} F_j(b_j)
\end{equation}
 
The function $F_j(b_j)$, calculates the total gain or loss of allocating a package with value $b_j \in D$ to the node $s_j$. By this formulation, we can easily use dynamic programming to select the optimal value of each $b_j$ from the set $D$ considering the constraint that $\sum_{j=1}^{n} b_j \leq k$. Furthermore, there is no need to explore the whole values of $D$ for each $b_j$, since, the value of function $F_j$ would not change for most of the values in $D$. In this way, we can largely reduce the search space. To clear the explanation, consider the following example.

Consider that we are trying to find the best response for the player that his budget is $k=3$, and his set of available packages is $D=\{1,2,3\}$. Also, assume that $n=4$ ($n=|S|$, number of the influential nodes identified in the first phase of our framework). The given mixed startegy is $q=\{0.8:a^1,0.2:a^2\}$ (with probability 0.8, $a^1$ would be played, and with probability 0.2, $a^2$ would be played). And, $a^1=\{2,0,1,0\}$, $a^2=\{2,1,0,0\}$. The estimated influential values of the nodes in $S=\{s_1, s_2, s_3, s_4\}$ are respectively $V=\{4,3,2,1\}$. So, the problem can perfectly be described in the following way:
\begin{displaymath}
  \begin{array}{rrccccccc}
    \text{nodes} \rightarrow  \:    && S   & \{ & s_1 & s_2 & s_3 & s_4 & \} \\
    \text{values} \rightarrow  \:   && V   & \{ & 4 & 3 & 2 & 1 & \} \\ \hline
    \multirow{2}{*}{$q \rightarrow \: $} & 0.8 :& a^1 & \{ & 2 & 0 & 1 & 0  & \} \\
    &0.2 :& a^2 & \{ & 2 & 1 & 0 & 0 & \} \\ \hline
    \text{best\_resp} \rightarrow \:    & & B   & \{ & b_1 & b_2 & b_3 & b_4 & \} \\
  \end{array}
\end{displaymath}
Here the gain of allocating a package with value 3 to $s_1$ is $F_1(3)=0.8 \times f(3,2)+0.2 \times f(3,2) = +4$. In the same way, $F_1(2)=0$ and $F_1(0) = F_1(1) = -4$. So the effect of $b_1=1$ and $b_1=0$ are the same, thus $b_1=1$ should not be explored as it is just wasting the budget. If we continue the calculations and use dynamic programming, we would get to $B=\{3,0,0,0\}$ and $\mathit{payoff}(B,q)=1.8$ (the maximum possible payoff).

\section{Experiments}
\label{sec:experiments}
In this section, through exhaustive experiments, we separately evaluate the performance of the payoff estimation method and the proposed framework.

The framework and experiments are implemented in python language using networkx library~\footnote{\url{https://networkx.github.io/}} (version 1.11), and Nash equilibrium in the restricted game (line 3 of double oracle algorithm, algorithm~\ref{alg:double_oracle}) is calculated using gambit framework~\footnote{\url{http://www.gambit-project.org/}} (version 16). Experiments are conducted on a machine with Intel Core i5 3.4 GHz 4 cores 64-bit processor and 32 GB memory. 

The source code of all the implementations are available in the project's repository~\footnote{\url{https://github.com/ah-ansari/cim}}.

\subsection{Experiment Setup}
\label{sec:experiments_settings}
Our experiments are conducted on the standard datasets used in the literature of influence maximization. Datasets and their statistics are listed in Table~\ref{tab:datasets}. Ca-HepTh is the collaborative network from Arxiv. Ego-Facebook consists of the circles from Facebook. Gnutella consists of the Gnutella peer-to-peer file-sharing network. Wiki-Vote represents who has voted whom in the administrator selection process in Wikipedia. Datasets are available in the Stanford Large Network Dataset Collection website~\footnote{\url{https://snap.stanford.edu/data/index.html}}.

For the competitive-IC model, the propagation probability of each edge $e$ is considered as  $P_e=1/\mathit{indegree}(v)$, where $v$ is the node that $e$ points to. This setting is widely adopted in previous studies~\cite{tang2014influence,tang2015influence,lin2015learning}. Further parameters of each evaluation are described in their own subsection.

\begin{table}
\caption{Overview of the datsets}
\label{tab:datasets}
\centering
\begin{tabular}{cccc}
\toprule
Name & Type & Node Number & Edge Number\\
\midrule
Ca-HepTh & Undirected & 9877 & 25998\\
Ego-Facebook & Undirected & 4039 & 88234\\
p2p-Gnutella08 & Directed & 6301 & 20777\\
Wiki-Vote & Directed & 7115 & 103689\\
\bottomrule
\end{tabular}
\end{table}

\subsection{Evaluation of the Influential Value Estimation Method}
The objective of the proposed heuristic method was to assign an influential value to the nodes in $S$ and then calculate the payoff based on this value. There exists no similar method in the literature that calculates the payoff in this way (by assigning a value to the nodes), so, we could only validate our method against the followings:
\begin{itemize}
  \item Simple RR-set based method: This is the method that we have extended. In this method, the influential value of each node $s_i \in S$ is calculated by estimating the influence spread of the singleton set $S'=\{s_i\}$ using the non-competitive RR-set based method (refer to Section~\ref{sec:non_competitive_rr_sets}): $\mathit{value}(s_i) = E[I(S')] = \frac{N}{\theta}C(S')=\frac{N}{\theta}C({s_i})$.
  \item Centrality-based heuristics: Centrality-based measures like Degree, Betweenness, Closeness, etc. can also be used to assign an influential value to the nodes in $S$. These methods are used to benchmark the effectivity of the proposed method and are all available in the networkx library.
\end{itemize}

The evaluation is conducted in this way that, first, the set $S$ with size $n=50$ is identified. Then, 20 experiments are executed for each dataset. In each experiment, the nodes in $S$ are randomly assigned to either $S_1$, or $S_2$, or none of the sets. This is to simulate different states that nodes in $S$ might take in the competition. After these random assignments, the actual value of $E[I(S_1)-I(S_2)]$ is calculated by the average of $I(S_1)-I(S_2)$ in 5000 rounds of repeated diffusion in the network according to the competitive-IC model. The absolute error would be the absolute difference between the actual value and the value estimated using the value-based methods. Finally, the mean absolute error of these 20 experiments for each method is reported in Table~\ref{tab:payoff_experiment}. 

According to the results, our method has significantly performed better than the SimpleRRset method; this proves that our extension is valid. Also, the heuristic centrality-based measures have all failed. An abnormal pattern in the results is the significant failure of the simpleRRset method under the p2p dataset (mean absolute error of 626.42). This is because this method does not consider the fact that the created RR sets may be covered by more than one node from the set $S$, and this effect is more considerable under the p2p dataset compared to other datasets.

\begin{table}
\caption{Mean absolute error of payoff estimation methods}
\label{tab:payoff_experiment}
\centering
\begin{tabular}{ccccc}
\toprule
Method & CA-HepTh & Facebook & p2p & Wiki-Vote\\
\midrule
\rowcolor{HColor} OurMethod & 6.33 & 15.91 & 83.76 & 9.13\\
SimpleRRsets & 20.27 & 55.70 & 626.42 & 23.71\\
Degree & 86.90 & 144.89 & 302.78 & 60.49\\
Betweenness & 86.85 & 144.90 & 302.78 & 60.70\\
Closeness & 86.18 & 144.61 & 302.20 & 60.13\\
Eigenvector & 86.89 & 145.34 & 302.74 & 60.69\\
Katz & 82.71 & 142.49 & 298.60 & 56.69\\
CoreNumber & 74.11 & 990.09 & 284.54 & 188.43\\
PageRank & 86.93 & 145.38 & 302.79 & 60.73\\
\bottomrule
\end{tabular}
\end{table}

\subsection{Evaluation of the Proposed Framework}
In this section, we conduct several experiments to show the effectivity of our proposed framework. We mainly focus on the symmetric case where $k_1=k_2=k$ and the advertising packages available to the players are $D_1=D_2=\{1,2,...,k\}$. This is because this case is more challenging. Winning the game in the other case, where one of the players has a larger budget, is not too difficult. Also, no similar framework exists in the literature to be compared with our method, and so, we have used the following strategies in our experiments:
\begin{enumerate}
  \item 1each: One unit of the budget is allocated to the $k$ nodes identified by TIM+.
  \item 2each: Two units of the budget are allocated to the $k/2$ nodes identified by TIM+.
  \item Random: Considering the $k$ nodes, this strategy starts from the first node and allocates a random value greater than zero to each node with the considerataion to keep $\sum_{i}{a_i}=k$.
  \item Random($z$): Like the above strategy, but it does not allocate a value greater than $z$ to a node (the maximum amount of budget allocated to a node is $z$).
\end{enumerate}
The 1each strategy resembles the studies that focused on the seed selection scenario, where the budget only identifies the seed set size. The Random strategy is an agent with no rationality, and Random($z$) is an improved version of Random.

As the experimental parameters, we have tested the values of $k= 10, 20, 30$. For $n$, the size of the set $S$, the values of $n= k, k+5, k+10$ are tested to see which parameter would perform better. Note that it is not wise to set $n < k$ since, in this way, the framework cannot investigate the case that the maximum possible number of nodes is selected. The double oracle algorithm is initialized (line 1 of Algorithm~\ref{alg:double_oracle}) with 1each strategy to make the results reproducible. Also, the complete convergence is used as the termination condition in $k=10,20$. However, in $k=30$, due to long running time, for all datasets except p2p, we have used $\epsilon=0.5$ and for p2p we have used $\epsilon=1$ as termination condition. In the results, $\mathit{CBA}(z)$ denotes our proposed framework, \textbf{C}ompetitive \textbf{B}udget \textbf{A}llocation (CBA) framework, where $n=z$.

\subsubsection{General Experiments}
As the first step, we benchmark the performance of our proposed framework (CBA), 1each, and 2each strategies by performing competitions between these strategies (as the first player) against the two Random, Random(3) strategies. Each competition is repeated 1000 times, and the win percentage of these methods is reported in Table~\ref{tab:benchmark}. Note that, in the Influence Maximization game, a win is when $I(S_1)-I(S_2)>0$. For each dataset and value of the budget, the method that has achieved the highest win percentage is highlighted. For example, in CA-HepTh dataset with $k=10$, 1each strategy has defeated Random strategy in 86.4\% of the competitions and has achieved the highest win percentage.

The Random(3) strategy is just the Random strategy improved by integrating a little bit of knowledge (to not allocate a value higher than 3). Note how the performance of the methods drops when tested against this strategy. In CA-HepTh dataset with $k=10$, the performance of 1each drops from 86.4\% to 54.4\%. Also, mostly the 1each and 2each strategies have achieved the highest win percentage in the competitions against Random. However, when we are testing against Random(3), our proposed framework achieves the highest win percentage. Another important point is that the effect of a non-perfect strategy, like the random strategy, gets bold when the value of budget increases, and this is why the win percentages increase as the value of $k$ increases.

\begin{table}
\caption{Win\% in benchmarking against Random and Random(3)}
\label{tab:benchmark}
\centering
\begin{tabular}{cccccccc}
\toprule
 & \multicolumn{3}{c}{vs Random} & \phantom{a} &\multicolumn{3}{c}{vs Random(3)}\\
\cmidrule{2-4} \cmidrule{6-8}
Method & k=10 & k=20 & k=30 && k=10 & k=20 & k=30\\
\midrule
\multicolumn{8}{c}{(a) CA-HepTh}\\
CBA(k)    & 77.1 & 97.4 & 99.8 && 51.6 & 56.2 & 63.1\\
CBA(k+5)  & 82.8 & 97.8 &\cellcolor{HColor} 99.9 && 56.7 & 60.4 &\cellcolor{HColor} 66.6\\
CBA(k+10) & 82.3 & 98.3 & 99.8 &&\cellcolor{HColor} 58.7 &\cellcolor{HColor} 60.7 & 66.4\\
1each     &\cellcolor{HColor} 86.0 &\cellcolor{HColor} 99.1 &\cellcolor{HColor} 99.9 && 54.9 & 57.0 & 65.5\\
2each     & 76.3 & 93.3 & 98.1 && 47.4 & 47.7 & 51.8\\
\multicolumn{8}{c}{(b) Facebook}\\
CBA(k)    & 57.0 & 70.0 &\cellcolor{HColor} 73.2 && 52.4 &\cellcolor{HColor} 68.0 & 73.9\\
CBA(k+5)  & 56.6 & 70.8 & 71.9 && 51.8 & 63.3 & 74.6\\
CBA(k+10) & 54.3 & 70.6 & 71.0 && 49.9 & 64.9 &\cellcolor{HColor} 76.6\\
1each     & 50.6 & 54.3 & 59.1 && 13.7 & 8.1 & 5.7\\
2each     &\cellcolor{HColor} 78.2 &\cellcolor{HColor} 72.0 & 67.3 &&\cellcolor{HColor} 58.9 & 51.9 & 48.5\\
\multicolumn{8}{c}{(c) p2p}\\
CBA(k)    & 80.3 & 93.1 & 99.4 && 52.1 & 50.7 & 55.3\\
CBA(k+5)  & 80.5 & 97.0 & 99.5 &&\cellcolor{HColor} 56.9 & 58.5 & 59.9\\
CBA(k+10) & 81.3 & 96.9 & 99.8 && 55.7 &\cellcolor{HColor} 58.6 &\cellcolor{HColor} 63.0\\
1each     &\cellcolor{HColor} 83.4 &\cellcolor{HColor} 99.3 &\cellcolor{HColor} 100.0 && 49.5 & 49.0 & 46.9\\
2each     & 73.5 & 91.4 & 99.3 && 49.8 & 55.5 & 49.4\\
\multicolumn{8}{c}{(d) Wiki-Vote}\\
CBA(k)    & 68.9 & 93.8 & 98.6 && 51.0 &\cellcolor{HColor} 58.6 &\cellcolor{HColor} 63.7\\
CBA(k+5)  & 72.0 & 94.4 & 99.0 && 52.3 & 58.5 & 60.1\\
CBA(k+10) & 73.4 & 93.5 & 99.0 && 50.2 & 57.5 & 62.4\\
1each     &\cellcolor{HColor} 77.2 &\cellcolor{HColor} 97.9 &\cellcolor{HColor} 99.9 && 40.8 & 32.9 & 40.2\\
2each     & 74.9 & 90.3 & 96.5 &&\cellcolor{HColor} 53.6 & 48.3 & 49.7\\
\bottomrule
\end{tabular}
\end{table}

In the next experiment, we conduct competitions between our method and 1each, 2each strategies. The average (avg.) and standard deviation (std.) of the $I(S_1)-I(S_2)$ in 1000 repeated competitions alongside with the win percentage are reported in Table~\ref{tab:competition_pairs}. For our framework, we have only reported CBA($k+5$) as the value of $n=k+5$
has performed slightly better than other values of $n$.
According to the results, in some cases, our method has won the game ($win\% \gg  50\%$, and also $avg. \gg 0$). In other cases, the results of the competition are a draw, $win\% \simeq 50\%$, and the value of avg. is small comparing to the value of std. However, our proposed framework has not lost any of these competitions with a large margin. Note that there is a correlation between win\% and avg; the value of avg. increases when win\% increases.

Another important point from Table~\ref{tab:competition_pairs} is the high value of the standard deviations. This is caused by two sources of randomness. First, the diffusion model is very stochastic. Second, the first player is playing the nash strategy, which is a mixed strategy, and so the actions are not deterministic and selected according to a probability distribution.

\begin{table}
\caption{Our method against 1each, 2each}
\label{tab:competition_pairs}
\centering
\begin{tabular}{cccc@{}ccccc@{}ccc}
\toprule
& \multicolumn{3}{c}{k=10} &\phantom{a}& \multicolumn{3}{c}{k=20} &\phantom{a}& \multicolumn{3}{c}{k=30}\\
\cmidrule{2-4} \cmidrule{6-8} \cmidrule{10-12}
dataset & win\% & avg. & std. && win\% & avg. & std. && win\% & avg. & std.\\
\midrule
\multicolumn{12}{c}{CBA(k+5) against 1each}\\
CA-HepTh   & 49.8 & 1.0 & 123.8  && 49.1 & 0.8 & 138.0   && 50.2 & 4.0 & 143.2 \\
Facebook   & 64.2 & 89.0 & 194.3 && 77.2 & 164.5 & 203.4 && 79.1 & 152.8 & 195.9 \\
p2p        & 53.6 & 20.4 & 504.4 && 54.2 & 49.3 & 432.7  && 47.3 & -14.8 & 350.8 \\
Wiki-Vote  & 50.4 & 1.5 & 75.8   && 47.4 & -3.4 & 79.0   && 46.3 & -4.4 & 78.6 \\
\multicolumn{12}{c}{CBA(k+5) against 2each}\\
CA-HepTh   & 49.4 & 4.3 & 101.7   && 50.9 & 0.2 & 115.5   && 47.5 & -7.7 & 124.3 \\
Facebook   & 45.5 & -5.7 & 185.3  && 63.8 & 83.9 & 193.2  && 75.7 & 160.1 & 213.5 \\
p2p        & 47.6 & -26.3 & 442.6 && 44.9 & -51.3 & 416.0 && 51.0 & 6.3 & 384.8 \\
Wiki-Vote  & 49.7 & 0.4 & 64.9    && 52.3 & 4.2 & 72.6    && 50.3 & 1.8 & 76.6 \\
\bottomrule
\end{tabular}
\end{table}

As the last experiment in this subsection, we conduct competitions between our method against itself with different values of $n$. This is mainly to identify which value of $n$ performs better. Results are reported in Table~\ref{tab:cba_diff_n}. According to the results, none of the parameters could significantly defeat others. Therefore, it seems that there is no significant difference between different values of $n$ tested; only $n=k+5$ has slightly performed better.

The key results from the experiments conducted so far are: 1. The performance of our method was better than the 1each/2each strategies in benchmarking against Random(3). 2. We have not lost any competition against other strategies. 3. The hint that the result of $n=k+5$ seems to be marginally better than other values of $n$. In the next subsection, we more accurately demonstrate the superiority of our framework compared to other strategies.

\begin{table}
\caption{Our method against itself with different values of $n$}
\label{tab:cba_diff_n}
\centering
\begin{tabular}{ccccc@{}cccc@{}ccc}
\toprule
 &\multicolumn{3}{c}{$(n=\!k\!+\!5)$ - $(n=\!k\!)$}&\phantom{a}&\multicolumn{3}{c}{$(n=\!k\!+\!5)$ - $(n=\!k\!+\!5)$}&\phantom{a}&\multicolumn{3}{c}{$(n=\!k\!+\!5)$ - $(n=\!k\!+\!10)$}\\
\cmidrule{2-4} \cmidrule{6-8} \cmidrule{10-12}
dataset & win\% & avg. & std. && win\% & avg. & std. && win\% & avg. & std.\\
\midrule
\multicolumn{12}{c}{$k=10$}\\
CA-HepTh   & 51.3 & 10.4 & 109.9 && 49.7 & -0.7 & 118.5 && 47.8 & -3.3 & 111.6 \\
Facebook   & 51.0 & -5.5 & 201.2 && 49.1 & -3.4 & 204.0 && 51.7 & 0.7 & 209.6 \\
p2p        & 51.7 & 20.1 & 489.7 && 50.8 & 17.0 & 531.6 && 49.6 & -2.3 & 543.7 \\
Wiki-Vote  & 47.6 & -1.7 & 70.5  && 50.7 & -0.9 & 74.1  && 53.0 & 3.9 & 69.4 \\
\multicolumn{12}{c}{$k=20$}\\
CA-HepTh   & 51.3 & 7.6 & 122.3 && 49.7 & -5.0 & 133.5 && 48.8 & -1.7 & 130.2 \\
Facebook   & 52.7 & 7.7 & 207.2 && 51.5 & 5.6 & 221.2 && 52.4 & 5.1 & 212.4 \\
p2p        & 53.4 & 24.4 & 406.3 && 49.5 & -4.1 & 438.6 && 50.4 & 6.8 & 466.4 \\
Wiki-Vote  & 49.4 & 0.7 & 80.3 && 49.5 & 2.3 & 80.9 && 46.7 & -3.9 & 79.7 \\
\multicolumn{12}{c}{$k=30$}\\
CA-HepTh   & 48.7 & -4.8 & 140.2 && 51.7 & 2.9 & 138.3 && 50.3 & 0.0 & 134.8 \\
Facebook   & 49.8 & -2.2 & 210.1 && 52.7 & 13.6 & 209.1 && 48.5 & 0.2 & 209.4 \\
p2p        & 51.4 & 13.9 & 409.8 && 49.6 & 1.8 & 418.8 && 50.2 & 3.3 & 416.1 \\
Wiki-Vote  & 50.1 & 4.1 & 79.5 && 48.5 & 2.8 & 80.6 && 49.0 & -0.4 & 79.8 \\
\bottomrule
\end{tabular}
\end{table}

\subsubsection{Why should one company adopt our framework?}
From a game-theoretic point of view, in the symmetric case, which is the case of these experiments, the Nash equilibrium (proposed framework) guarantees not to be the absolute loser of any competition (refer to Section~\ref{sec:framework}). However, for any non-nash strategy, there exists at least one strategy that can definitely defeat it. This is the main superiority of our framework.

To illustrate the point above, we formed competitions between 1each, 2each, Random(2), and Random(3) strategies against their pure best responses calculated by our best response oracle. Our method is also evaluated against these best responses and its own best response. Like before, each competition is repeated 1000 times, and results are reported in Table~\ref{tab:best_responses}.
The results of this experiment demonstrate how badly these strategies have lost the game to their best responses. In the Facebook dataset, for example, none of these methods could win any competition, win\% of 0. However, our method's performance against these best responses and its own best response is completely following what was guaranteed.

\begin{table}
\caption{Competitions between methods against their best responses, our framework is also tested}
\label{tab:best_responses}
\centering
\begin{tabular}{@{}cccc@{}ccccc@{}ccc}
\toprule
& \multicolumn{3}{c}{k=10} &\phantom{a}& \multicolumn{3}{c}{k=20} &\phantom{a}& \multicolumn{3}{c}{k=30}\\
\cmidrule{2-4} \cmidrule{6-8} \cmidrule{10-12}
dataset & win\% & avg. & std. && win\% & avg. & std. && win\% & avg. & std.\\
\midrule
\multicolumn{12}{c}{1each against BestResponse(1each)}\\
CA-HepTh   & 34.5 & -33.0 & 93.9   && 24.0 & -71.2 & 99.1  && 20.6 & -82.2 & 108.7 \\
Facebook   & 0 & -506.3 & 92.0   && 0 & -642.8 & 110.3 && 0 & -658.9 & 101.4 \\ 
p2p        & 24.1 & -246.9 & 370.1 && 8.5 & -414.1 & 296.4 && 2.7 & -498.6 & 262.0 \\
Wiki-Vote  & 6.9 & -67.9 & 45.2    && 1.9 & -106.9 & 48.9  && 0.4 & 118.0 & 46.0 \\
\multicolumn{12}{c}{CBA(25) against BestResponse(1each)}\\
CA-HepTh   & 49.4 & 4.3 & 101.7   && 50.9 & 0.2 & 115.5   && 47.5 & -7.7 & 124.3 \\
Facebook   & 45.5 & -5.7 & 185.3  && 63.8 & 83.9 & 193.2  && 75.7 & 160.1 & 213.5 \\
p2p        & 47.6 & -26.3 & 442.6 && 44.9 & -51.3 & 416.0 && 51.0 & 6.3 & 384.8 \\
Wiki-Vote  & 49.7 & 0.4 & 64.9    && 52.3 & 4.2 & 72.6    && 50.3 & 1.8 & 76.6 \\
\multicolumn{12}{c}{2each against BestResponse(2each)}\\
CA-HepTh   & 20.9 & -46.7 & 71.3   && 7.1 & -127.5 & 93.4  && 2.1 & -195.4 & 96.4 \\
Facebook   & 0 & -271.3 & 88.6     && 0 & -635.8 & 102.6   && 0 & -751.0 & 98.4 \\
p2p        & 15.4 & -372.4 & 387.4 && 2.9 & -726.1 & 375.4 && 0.5 & -868.5 & 304.9 \\
Wiki-Vote  & 4.2 & -67.5 & 41.2    && 0   & -143.1 & 53.6  && 0 & -193.9 & 46.7 \\
\multicolumn{12}{c}{CBA(25) against BestResponse(2each)}\\
CA-HepTh   & 81.2 & 79.1 & 94.4   && 72.1 & 60.9 & 104.5  && 80.3 & 90.4 & 112.1 \\
Facebook   & 61.3 & 66.3 & 201.8  && 59.7 & 19.1 & 202.7  && 60.5 & 63.7 & 188.7 \\ 
p2p        & 80.6 & 326.7 & 383.9 && 61.8 & 124.7 & 375.0 && 57.2 & 74.6 & 351.1 \\
Wiki-Vote  & 63.1 & 27.8  & 66.9  && 53.5 & 10.7 & 71.5   && 66.6 & 30.5 & 67.7 \\
\multicolumn{12}{c}{Random(2) against BestResponse(Random(2))}\\
CA-HepTh   & 40.1 & -30.3 & 111.7  && 32.2 & -50.3 & 122.8  && 32.5 & -58.3 & 129.8 \\
Facebook   & 9.0 & -253.8 & 177.9  && 0 & -563.4 & 109.7    && 0 & -625.8 & 106.0 \\
p2p        & 32.4 & -215.5 & 501.1 && 28.4 & -241.8 & 456.2 && 16.0 & -403.0 & 423.7 \\
Wiki-Vote  & 31.4 & -40.7 & 80.9   && 11.5 & -80.5 & 67.5   && 8.7 & -93.3 & 69.1 \\
\multicolumn{12}{c}{CBA(25) against BestResponse(Random(2))}\\
CA-HepTh   & 51.0 & 5.3 & 113.6   && 48.9 & -2.5 & 131.1  && 47.7 & -3.6 & 135.9 \\
Facebook   & 49.0 & 3.1 & 181.5   && 57.9 & 14.5 & 205.1  && 57.1 & 40.7 & 184.0 \\
p2p        & 47.1 & -33.9 & 491.2 && 45.9 & -31.4 & 440.9 && 55.2 & 51.7 & 393.6 \\
Wiki-Vote  & 52.6 & 3.3 & 70.4    && 52.1 & 0.8 & 75.0    && 49.0 & -1.9 & 77.7 \\
\multicolumn{12}{c}{Random(3) against BestResponse(Random(3))}\\
CA-HepTh   & 33.5 & -39.8 & 102.6  && 29.6 & -62.4 & 113.8  && 15.4 & -123.9 & 125.3 \\
Facebook   & 32.5 & -102.5 & 231.8 && 0 & -483.3 & 102.4    && 0 & -582.7 & 116.8 \\
p2p        & 33.1 & -236.1 & 474.1 && 22.5 & -306.9 & 410.8 && 17.6 & -337.6 & 363.5 \\
Wiki-Vote  & 33.5 & -27.6 & 66.7   && 13.7 & -73.8 & 70.9   && 7.9 & -103.8 & 72.1 \\
\multicolumn{12}{c}{CBA(25) against BestResponse(Random(3))}\\
CA-HepTh   & 57.1 & 17.3 & 105.1 && 59.9 & 30.6 & 122.4 && 60.1 & 30.8 & 127.7 \\
Facebook   & 47.8 & -0.2 & 172.9 && 52.4 & 4.8 & 207.3  && 54.3 & 16.0 & 195.3 \\
p2p        & 54.6 & 43.0 & 464.6 && 57.6 & 75.5 & 414.5 && 51.5 & -0.5 & 393.4 \\
Wiki-Vote  & 49.1 & -0.9 & 71.9  && 50.9 & 4.4 & 78.0   && 49.7 & 4.4 & 71.4 \\
\bottomrule
\end{tabular}
\end{table}

\subsubsection{Running Time and Number of iterations to converge}
Finally, we report the running time and number of iterations it takes for the double oracle algorithm to converge in Table~\ref{tab:n_converge}. Only the worst case of $n=\{k, k+5, k+10\}$ for each value of $k$ is reported. The cardinality of the action space is $\binom{k+n-1}{n-1}$; this is why we observe a significant increase in running time as we increase the value of $k$.

Another important point is that the size of the graph does not affect the running time of the double oracle algorithm since we are solving the problem from a higher level of abstraction where the competition is formed around only the nodes in $S$. It is the distribution of the influential value of the nodes that affect the complexity of the game and the running time. For example, in k=30, the most significant running time was under the p2p dataset (with even a looser termination condition $\epsilon=1$), which is not our largest dataset.

\begin{table}
\caption{Running time and number of iterations to converge}
\label{tab:n_converge}
\centering
\begin{tabular}{cccc@{}cccc@{}cccc}
\toprule
& \multicolumn{3}{c}{k=10} &\phantom{a}& \multicolumn{3}{c}{k=20} &\phantom{a}& \multicolumn{3}{c}{k=30}\\
\cmidrule{2-4} \cmidrule{6-8} \cmidrule{10-12}
dataset & iter. & time & condition && iter. & time & condition && iter. & time & condition \\
\midrule
CA-HepTh   & 51 & 43s  & converge && 150 & 6h & converge && 162 & 17h & $\epsilon=0.5$\\
Facebook   & 45 & 47s  & converge && 101 & 0.9h & converge && 155 & 16h & $\epsilon=0.5$\\
p2p        & 69 & 345s & converge && 132 & 5h & converge && 184 & 47h & $\epsilon=1$\\
Wiki-Vote  & 72 & 220s & converge && 171 & 11h & converge && 150 & 8h & $\epsilon=0.5$\\
\bottomrule
\end{tabular}
\end{table}

\section{Conclusion and Future Work}
\label{sec:conclusion}
In this paper, we have presented the Competitive Influence Maximization problem from a higher level of abstraction, where parties first identify the influential nodes of the network. Then, they compete over these nodes by their advertising packages. Also, we have considered the action space to be discrete. This consideration makes a considerable difference to previous studies and is more realistic.  We propose an efficient framework with a novel payoff estimation method that calculates the Nash equilibrium. Our framework targets the case of two players and is entirely applicable when the players have different amounts of budget, and when their available advertising packages are different. Through several experiments, we evaluate each aspect of our work. Especially we show that any non-nash strategy would badly lose the game to its best response. However, this would not happen for our proposed framework, and it guarantees not to lose any competition in the symmetric case. This is the critical superiority of our method. 

Our framework is mainly built upon the TIM+ algorithm. However, any other RR-set based method like IMM can easily be mapped to our framework. Also, since the concept of RR-sets applies to the Linear Threshold and Trigerring based models, our payoff method can be extended to these diffusion models.

As future works, we first intend to extend the framework to more than two players. This will not be easy since, for example, the Minimax theorem does not hold in that case. Also, so far, we could not provide a theoretical approximation guarantee for our payoff estimation method, and further work is required in that direction. Finally, analyzing other extensions that make the problem more realistic, like considering the tendency of influential customers, can be considered as future work.

\bibliographystyle{unsrt}  
%\bibliography{mybibliography}  %%% Remove comment to use the external .bib file (using bibtex).


\end{document}